# 'Um, er': How meaning varies between speech and its typed transcript

## Harry Collins, Willow Leonard-Clarke and Hannah O'Mahoney[1]

## Introduction

Human societies communicate mainly with language. Most linguistic interactions are spoken but there are also various symbol systems: there is Morse code and other codes; there is semaphore with flags; there are various kinds of analogue codes such as the squiggles inscribed on a wax surface by a stylus driven by the changes in air pressure which constitute sound and the many electrically driven equivalents that make telephone and wireless a possibility. Here, however, we are concerned with the relationship between speech and the writing which is taken to represent it. Writing, of course, is not mainly used to encode speech; mostly written texts are communicative devices *sui generis* but here we are concerned only with writing that is intended to represent words that were initially created in the form of talk.

Sometimes writing that is intended to represent talk is not taken from talk. For example, novelists writing speech for their characters must set out plausible passages of talk first using the symbols which are conventionally used for written language. Presumably, some fiction writing schools teach novice writers how to do this so that the supposed speech looks plausible or, perhaps, representing it as spoken in an accent obviously associated with a certain class or region or country. This is sometimes used to indicate more or less intelligence, more or less courage, more or less privilege, more or less trustworthiness or more or less in the way of good and bad intentions. Sometimes writing comes before speech as when playwrights invent speeches for their characters which are later spoken on stage or film set. In that case the plausibility of the speech is 'experimentally' tested and there must be a lore about the way actors licence themselves to make small changes in a script – to interpret it – in such a way as to make less than optimum written speech more plausible. Another rather strange example is when academics write papers and then read them out at seminars or conferences: how what is heard relates to what would be understood if read is a curious but, perhaps, not very interesting problem, which results from the different lexical density of the two media – spoken and written discourse (Halliday, 1985). The case we deal with here is when writing comes after speech and is intended as a written record of what was said – a symbolically inscribed equivalent of the gramophone record or tape-recording and its digital equivalents. In such instances the idea is that the reader can recapture what was said by reading inscribed symbols just as they could recapture it by replaying and listening to some kind of electrical recording.

[1] Collins is Distinguished Research Professor at Cardiff University and conceived and designed the project which emerged from his gravitational-wave studies funded by the British Economic and Social Research Council and his Imitation Game studies funded by the European Research Council: ESRC (RES-000-22-2384) £48,698 `The Sociology of Discovery' (2007-2009); ESRC (R000239414) £177,718 `Founding a New Astronomy' **(2002-2006);** ESRC (R000236826) £140,000 `Physics in Transition' **(1996-2001)**. Advanced Grant and a Proof of Concept grant from the European Research Council: (269463 IMGAME) €2,260,083 `*A new method for cross-cultural and cross-temporal comparison of societies';* (297467 IMCOM) €150,000 `IMGAME Commercial*'*. Leonard-Clarke is an undergraduate in the School of Social Sciences at Cardiff University and carried out most of the fieldwork, funded from Collins's large ERC grant. O'Mahoney is a past research associate on the IMGAME project, and is mainly responsible for setting the results in the context of the range of approaches to the use of interviews across the social sciences.

Sometimes the equivalence or non-equivalence of the written record of speech and the speech which it is taken to represent is of vital importance[2]. In some legal settings, where judgements are based on written transcriptions of evidence, it could be a matter of life and death. One famous case that at least indicates the nature of the problem is the execution, in Britain, in 1953, of Derek Bentley for the murder of a police constable in which it was acknowledged that the person who pulled the trigger was his accomplice, Christopher Craig. Craig was too young to be executed. The key words spoken by Bentley were 'let him have it' which could be interpreted as the English vernacular for 'hit him' or 'shoot him' or as an instruction to Craig to hand over the gun to the policeman who was asking for it ('asking for it' is itself an ambiguous phrase). In this case the only form of representation that might have saved Bentley's life would have been an electrical recording that may have captured the nuance of what was said on the night. All other renderings of the phrase were open to the interpretation that led the jury at the trial to find Bentley guilty based on the prosecution's interpretation. The judge described Bentley as "mentally aiding the murder of Police Constable Sidney Miles" (Wikipedia, n.d.: Bentley's case was quashed in 1998). We do not know the extent to which the Bentley case turned on written transcripts of what Bentley said though we do know that only the policemen present at the time were able to interpret the lived speech. Thus, though it may have been that one or more of those policemen might have tried to mimic the intonation used, pretty well everyone only knows what was said from written sources. More recently, Heffer's studies have shown that the ways court depositions are transcribed can affect the way legal cases unfold.[3]

---

[2] Though it has been argued that there can never be be equivalence between spoken and written discourse (Catford, 1965).

We believe the general meaning-relationship of different ways of representing 'the same' words is a new subject for investigation which, as far as we know, has not been thought about in this broad way. The subject is the comparative study of typical interpretations of 'the same' passages of language when spoken and written and transmitted in other symbolic forms. Let's call this subject 'language code analysis', or LCA. It will extend to other means of encoding of language such as Morse Code, or the modern coding used for 'texting' or twitter. Of course, a major topic of LCA would be what is 'the same' when spoken and written or otherwise encoded, but this should not be taken as a problem that is bound to vitiate the enterprise. What happened to Derek Bentley shows that answers are going to be given to these questions which are very significant to some people and it would be wrong to turn away because the problem is a difficult one for the social sciences to resolve with certainty.

In this paper we present a small study that could be said to belong to the new subject of LCA. Its concern is with the transcription of interviews. The example used is taken from studies of science. The question asked is about how 'the same' passage of speech is interpreted when heard live and when transcribed according to certain conventions. Many contemporary approaches to text stress the indefinite interpretability of all text and, indeed, spoken utterances. We agree that, in a sense, 'the author is dead' and that anything can be interpreted as anything just as pictures can be seen 'in the fire' or shapes seen in a cloud. But what we do here is concentrate on the other side of the equation which for convenience's sake we will refer to as the 'affordance' of texts and spoken utterances. That anything can be interpreted as anything may be true, but it is not true that anything can *just as easily* be interpreted as one thing rather than another. The comparative difficulty of generating different interpretations reflects what we are calling their differential affordances. Without these differential affordances the very idea of a code or symbol system would make no sense.[4]

---

[3] Private communication
[4] Collins, 2017 forthcoming – *Artifictional Experts* – analysis the issue in terms of 'bottom-up' and 'top-down' interpretations of symbols.

**Background**

About 20 years ago, as part of his long-running project on the sociology of gravitational wave detection, Collins conducted one of many interviews with a gravitational-wave physicist. Collins selected an interview extract he had transcribed for publication and sent it to the respondent for checking. The respondent asked that the 'ums' and 'ers' be removed since he thought they gave the impression of more uncertainty than he intended. At first Collins demurred, saying that if he removed the hesitancies the extract would no longer be faithful to what the respondent had said. But later, it struck Collins that, when seen in textual form, the uhms and ers gave a greater impression of uncertainty when presented in text form than they did when spoken. When spoken they tended to be heard as place-holders while the speaker thought *carefully* about what was being said rather than indications that the speaker was unsure; when read from the page they indicated doubt. In this case it seems that a transcription that is to be faithful to what the speaker intended to convey has to be less than faithful to the stream of phonemes uttered.[5]

It also became clear that similar liberties had to be taken where speakers were not native English speakers. Thus a major part of the book that would eventually be written (Collins, 2004) concerns a bitter conflict between certain American physicists and certain Italian physicists. This conflict was illustrated by quotations from each group. But if transcribed faithfully the American group's claims would appear in near perfect English whereas the Italian group's claims would appear in broken English reducing the impression of authority. For this reason Collins adopted a policy of correcting the English of Italian respondents as it appeared on the page.

Twenty years after the initial realisation, when further consideration of the 'right way' to transcribe was occasioned by an Imitation Game project funded by the European Research Council, the issue arose again. Part of the project involved transcribing the conversations of groups of people as they played the game and Collins noticed that researchers were transcribing these interactions in great and pedantic detail and asked why they were bothering since, in Collins view, the aim of transcription at this stage was to identify the intended meanings of interlocutors rather than their speech patterns or the intricacies of their interactions. What seems to have happened is that accuracy in preparing transcripts seems to have more and more become the norm for many social scientists.[6] It was a version of this norm that accounts for Collins's initial reluctance to removeuhms and ers when it was requested by the respondent. These considerations led to the decision to test the difference in perceived meaning of hesitancies in textual representations of speech and in audio recordings. In other words, we sought to explore perceptions of spoken and written versions of what we take to be the same message. The results should help to answer the question of exactly how to transcribe for different purposes and to make a contribution to language code analysis in general, exemplifying one possible method for empirical research.

---

[5] Collins removed the um and ers and adopted this as his practice from then on.

**Method**

Having sought permission from the original respondent for this unanticipated use of the 20-year-old interview, we selected two short extracts from it, each heavily laden with uhms and ers. We transcribed these 'faithfully', keeping the uhms and ers, but without using the technical apparatus of conversational analysis as we wanted them to be easily readable by non-specialists; we did what might be referred to loosely as an orthographic 'accurate transcript' of each selection. The two transcripts are shown below as presented to participants, though each turn was separated by a blank line.

> **Extract 1**
> **Interviewee:** I believe, and [anonymised name] believes, that these gravity wave events that we're seeing are, should be so, infrequent,
> **Interviewer:** mm
> **Interviewee:** that, err, we should not try to have lots and lots of coincidences and look for little changes in it we have to, and, we,
> **Interviewer:** mm mm [overlap]
> **Interviewee:** we have to just look for, err, do the standard bayesian statistics, mm, and, err, uhh, look in, in that way
> **Interviewer:** Right
> **Interviewee:** Uhm, and, err, I m-an-an-and that's never going to change

---

[6] The development of transcription conventions in the social sciences has, in part, reflected the recognition that these choices should be of concern to social scientists. Gail Jefferson's conventions (Sacks et al., 1974; Jefferson, 2004), or variations thereof, are perhaps most commonly employed, to the point that they are regarded as *the* 'default transcription system' (Ayaß, 2015). Data transcribed in the strict Jeffersonian system is highly detailed, providing the reader with access to a wealth of information besides the actual words spoken by interlocutors, such as intonation, latching, breath sounds, timed pauses, etc. Some scholars advocate the use of a sort of 'Jeffersonian lite' transcription (Potter and Wetherell, 1987; Potter and Hepburn, 2005), in which less detail is provided but some insight into the non-linguistic characteristics of spoken interactions are reproduced in transcripts.

| |
|---|
| **Interviewer:** mm [overlap]<br>**Interviewee:** because it's just a different way of looking at the physics we do,<br>**Interviewer:** yeah [overlap]<br>**Interviewee:** and we have, we just have to understand that we as individuals are different<br>**Interviewer:** Yeah |
| **Extract 2**<br>**Interviewee:** And, err, I think three or four years ago that would have been totally ignored, was totally ignored,<br>**Interviewer:** mm [overlap]<br>**Interviewee:** I mean I know it was, uhh, the, err, theoretical physics community felt that our experience was absolutely, had nothing to do with, err, uh, with LIGO, LIGO was going to be so much better<br>**Interviewer:** mm [overlap]<br>**Interviewee:** we were just going to dry up and blow away on the wind.<br>**Interviewer:** mm mm, mm mm [overlap]<br>**Interviewee:** and err, uh now we are seeing, that, err, uhh, uhh, this experience is useful, more were that these detectors<br>**Interviewer:** mm [overlap]<br>**Interviewee:** actually may be useful as far as being able to, err, uhm, to, er, uh, develop, uh, t-be an additional confirmation of, of detection<br>**Interviewer:** Yeah [overlap]<br>**Interviewee:** even give information that the, that the, err, detectors can't give<br>**Interviewer:** More directionality and so forth<br>**Interviewee:** Yeah, right. |

*Two extracts from an interview about gravitational wave physics*

The audio extracts that correspond with these transcripts can be found at www.cf.ac.uk XXXX..... Readers should feel free to use these materials to repeat and test our findings as well as to assure themselves that our results accord with their own, native English-speaker, hearing and reading of the extracts.

For the experiment, the original sound recordings of the extracts were made easily accessible on a tablet computer. We now had audio and text versions of two extracts which we will henceforward refer to as A(udio)1, T(ranscript)1, A2 and T2. We now presented these to respondents – each respondent seeing one audio and hearing one text extract and being asked to assess them according to seven criteria. We were concerned that responses might be affected by the order in which the extracts were presented and so we used an experimental design in which all four possible orderings were applied. The experiment was, therefore, conducted in 'sets' of four participants arranged as shown in **Error! Reference source not found.**:

|  | PARTICIPANTS | | | |
|---|---|---|---|---|
|  | **P1** | **P2** | **P3** | **P4** |
| **Scores first** | A1 | T1 | A2 | T2 |
| **Scores second** | T2 | A2 | T1 | A1 |

*Table 1: Distribution of material to participants*

Directly after hearing or reading each extract, participants were asked to 'score' various perceived features of the material. A compressed version of the scoring sheet presented to participants is shown in Table 1.

1) **The interviewee is friendly**
   *Strongly Disagree*   1   2   3   4   5   6   7   8   9   10   *Strongly Agree*
2) **How old do you think the interviewee is?**
   Under 25   26-35   36-45   46-55   56-65   65-75   Over 75
3) **What nationality do you think the interviewee is?**
   British   Italian   American   Australian
4) **The interviewee seems authoritative**
   *Strongly Disagree*   1   2   3   4   5   6   7   8   9   10   *Strongly Agree*
5) **The interviewee is certain about what they are saying**
   *Strongly Disagree*   1   2   3   4   5   6   7   8   9   10   *Strongly Agree*
6) **The interviewee works in a university**
   *Strongly Disagree*   1   2   3   4   5   6   7   8   9   10   *Strongly Agree*
7) **The interviewee knows more in the area being discussed than the interviewer**
   *Strongly Disagree*   1   2   3   4   5   6   7   8   9   10   *Strongly Agree*

*Table 1: Scoring sheets used to judge audio and transcribed extracts.*

As can be seen, question 5 concerns the degree of certainty expressed by the interviewee. The other questions were intended as distractions so as to prevent participants from trying to 'second-guess' the purpose of the exercise, and as comparisons so that we could see if the certainty effect would stand out from other effects. As it happens some of the other criteria were, understandably in retrospect, correlated with the certainty choice. Thus there was a rough correlation between perceived certainty and perceived authority and authority was also roughly correlated with perceived age. Nevertheless, the almost complete uniformity of responses to the certainty question did stand out and was in marked contrast to the pattern of perceived attributes such as friendliness and nationality. It is the certainty question upon which we concentrate our analysis with some other results presented for contrast.

Initially, 7 sets of four were completed giving 28 results in all. Table 2 shows how the certainty question was scored by the 28 participants. Each vertical line represents a single respondent. Here the black dots show the degree of certainty they attributed to the audio extract and the crosses indicate the certainty they attributed to the text extract.

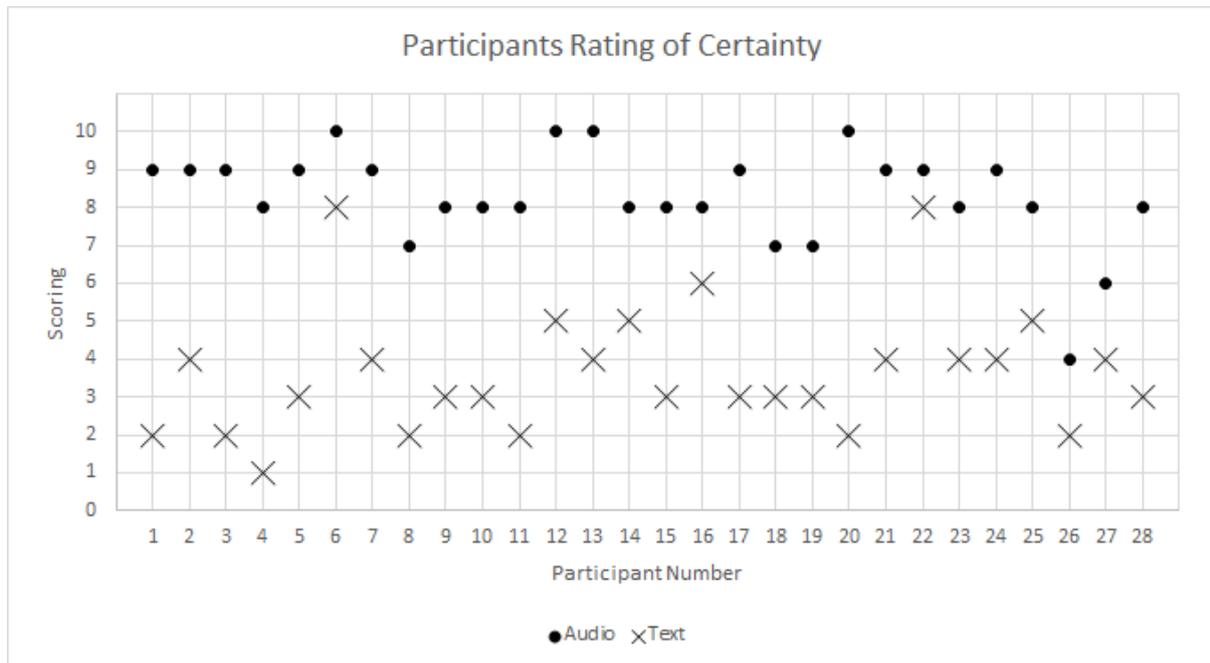

*Table 2: Scoring of the 'certainty' question*

As can be seen, every participant scored the certainty of the audio extract higher than the certainty of the text extract and the difference was strongly marked in nearly every case. Here the response is so uniform and the difference so marked that arithmetical or statistical analysis is otiose.[7] However, we did think it important to check that the order in which each extract was delivered did not have a substantial impact upon the result. Table 3 shows that, although there were differences, which were interestingly more marked in the transcribed than audio materials, they were small. It is worth noting the 2$^{nd}$ extract in both audio and transcribed form were judged to be less certain than the 1$^{st}$. Brennan and Williams (1995) found that the presence of pauses, 'ums' and 'uhs' in speech impacted negatively on the extent to which listeners felt that speakers were knowledgeable. This might account for the lower average certainty scores of both the audio and transcribed versions (A2, 7.86 and T2, 3.36) of the 2$^{nd}$ as compared to the 1$^{st}$ extract (A1, 8.71 and T1, 3.93), as this had twice as many disfluencies (including uhm, er, uhh, and mmm sounds).

---

[7] This is a case where we have good reasons to expect that there would be uniformity across the whole population of native English speakers so questions of representativeness of the sample are not appropriate – see Collins and Evans 2016 'Probes, surveys and the ontology of the social'. *Journal of Mixed Methods Research*, forthcoming

|  | A1 | T1 | A2 | T2 |
|---|---|---|---|---|
| Scores First | 8.86 | 4.71 | 7.86 | 3.29 |
| Scores Second | 8.57 | 3.14 | 7.86 | 3.43 |
| Mean | 8.71 | 3.93 | 7.86 | 3.36 |

| Overall means | A1 +A2 | T1+T2 |
|---|---|---|
|  | 8.29 | 3.64 |

*Table 3: Average 'certainty' scores of transcribed and audio material, disaggregated by the order in which they were delivered to participants.*

Table 4 illustrates a similar outcome from the question concerning the assumed authoritativeness of the interlocutors to that found in the responses to the 'certainty' question. Here again, the participants were almost unanimous in thinking that the audio extracts sounded more authoritative than the transcribed versions.

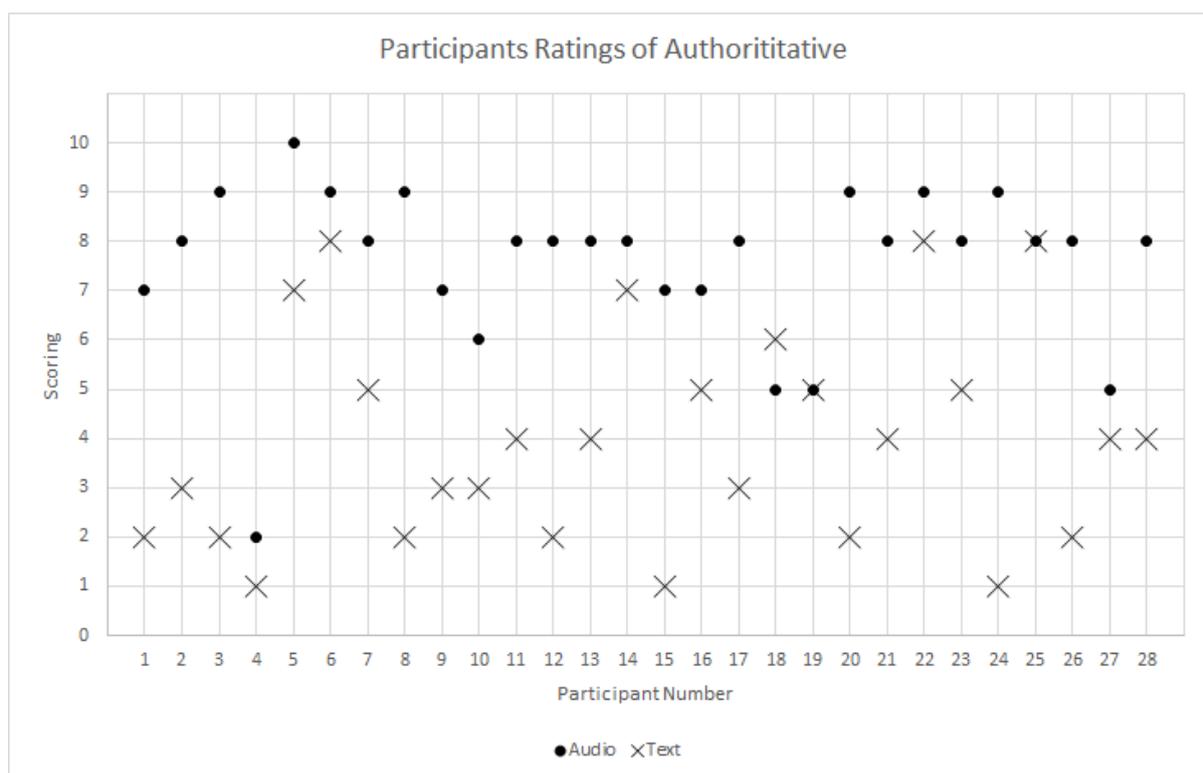

*Table 4: Scoring of the 'authoritative' question*

### *Sampling modifications: Recruiting an 'expert set'*

The sample for the initial experiment was largely one of convenience, and included university staff, undergraduate and postgraduate students, and professionals employed outside of higher education. Following the observation that the transcripts would probably be perceived very differently by those with expertise and experience in reading and/or working with these kinds of material, an 'expert' set were recruited comprising academics skilled in discourse or conversation analysis. The results from this set are reported in

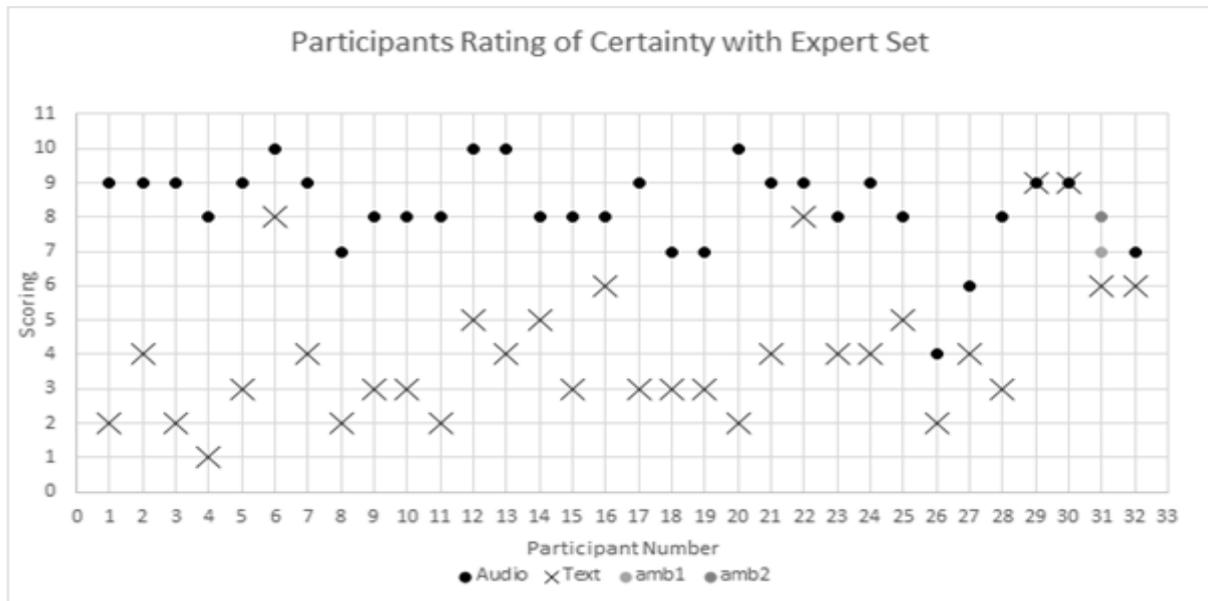

Table 5 and Table 6, with the expert set comprising participants numbered 29 to 32. The two small grey dots in column 31 represent a respondent who insisted on giving a score *between* 7 and 8. Whilst the audio files were still deemed more certain than the transcribed material (audio average: 8.125, transcript average: 7.5), and more authoritative (audio average: 9, transcript average: 8), these differences were markedly smaller than those found in the 'lay' experiment.

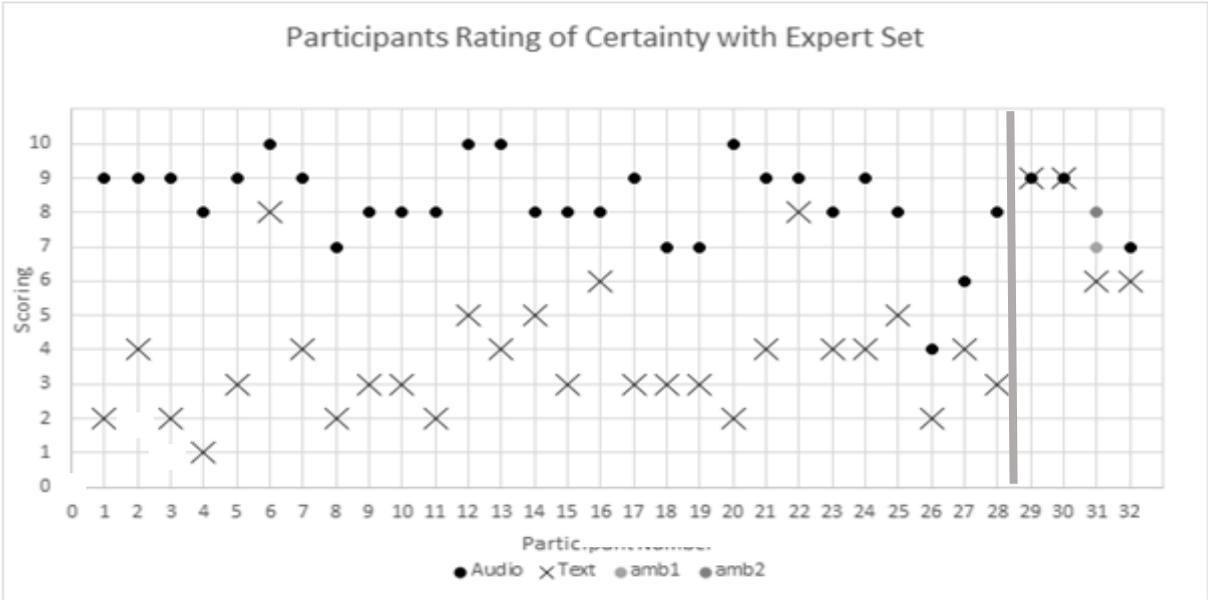

Table 5: Scoring of 'certainty' question with 'expert set' included

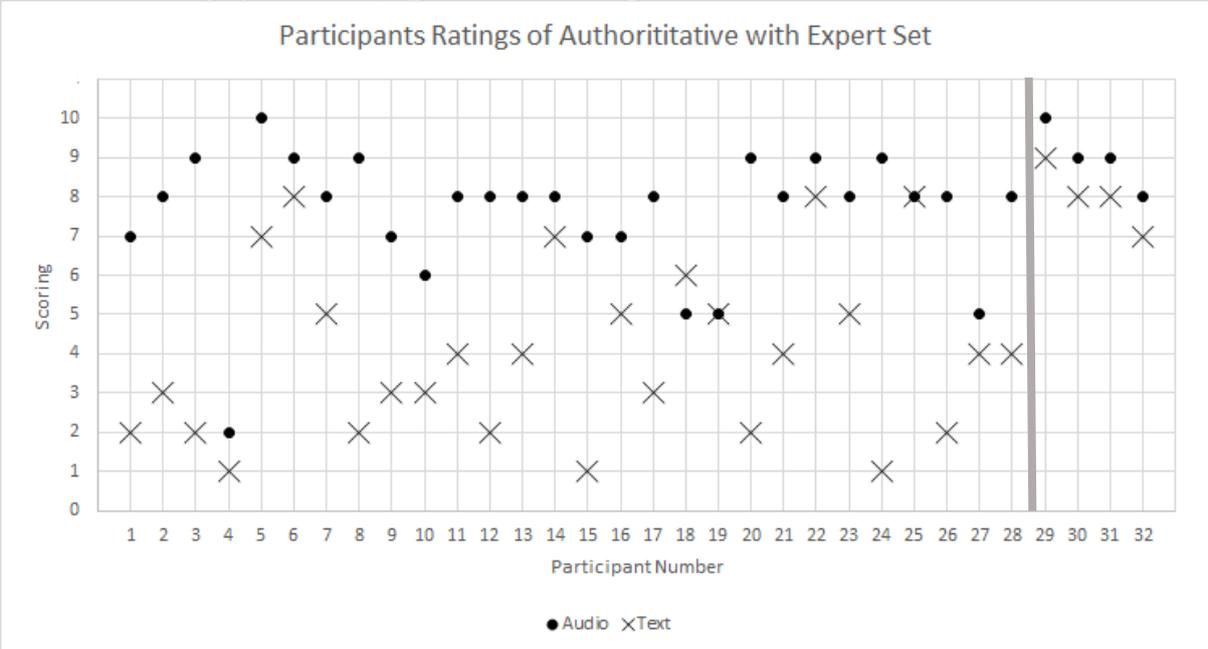

Table 6: Scoring of 'authoritative' question with 'expert set' included

To try to eliminate one more possible source of doubt about the significance of the result, we distributed the same material to a further four participants who were not experts in discourse analysis, but this time we removed all disfluencies from the transcribed material. In this mini-experiment, the average rating of the certainty of *both* the audio and transcribed material was exactly the same: 6.75. The distribution of these scores is illustrated in 8.

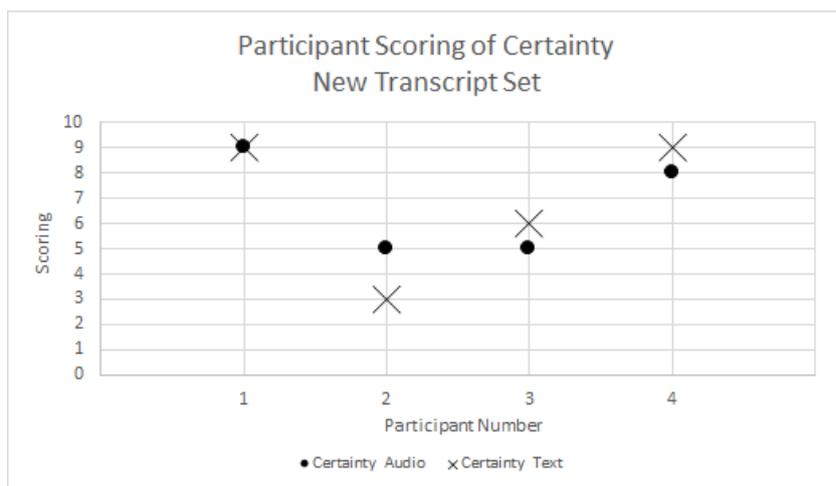

*Table 7: Scoring of 'certainty' question*

**Conclusion**

Using a simple investigative method we have shown that the same 'phonemes' have very different meanings when presented in spoken and written form. The difference is so striking as to render a statistical analysis otiose. We also show that those accustomed to dealing with transcripts are far less ready to agree that there is such a difference though this result is based on small numbers. We also show, again on the basis of small numbers, that the initial difference reported is the result of inclusion of the hesitancies since when these were removed from the transcript, the average score for the audio and text transcripts was almost identical in terms of the certainty conveyed, both scoring highly. The differences indicate strongly that much more consideration should be given to the conventions of transcription when it is used for different purposes. We note that sometimes the convention adopted can have grave consequences. We have also suggested that this research could be just the beginning of a new subject which we have called 'language code analysis' (LCA).

**Appendix 1: Use of transcription conventions in sociology**

Variations in the transcription of interview data can 'directly influence the nature and direction of the analysis' (Sandelowski, 1994: 311). These variations include the degree of meticulousness in verbatim transcription of speech as well as decisions about whether or not to preserve other oral but non-linguistic acts such as laughing, crying, sighing, gestures, and articulated disfluencies such as uhms and ers. As Silverman notes, in making decisions about what should and should not be included in a transcript, 'everything depends on the status which we accord to the data' (Silverman, 2011: 199), and whether the data is seen as a resource for analysis or itself a *topic* of analysis (Rapley, 2001; Seale, 1998). Transcription is not neutral, it is itself a form of decision making (Hepburn and Bolden, 2012; Mishler, 1991; Hammersley, 2010), and the choices we make relating to the presentation of our data should simultaneously be appropriate to the questions we seek to answer and the audience with whom we wish to communicate.

The use of 'full' Jeffersonian style transcripts is appropriate to the needs of, for example, conversation analysts, who explore the sequential and situated nature of interactions (Atkinson and Heritage, 1984). Conversation Analysis (or CA) entails a highly contextual and situated approach to understanding interactional data, and is related to the discipline of ethnomethodology insofar as both seek to understand the production of social order as manifested in human interaction (Roulston, 2006). However, are such conventions appropriate in research where the linguistic *performance* is secondary to the information contained within the speech act, or, indeed, wholly irrelevant to the research in question?

***Inclusion of 'ums' and 'ers' in transcribed material***

Even outside of CA, uhms and ers are often included in written transcripts. An argument *for* this practice has to do with the fact that these utterances are not simply 'disfluencies' or 'filled pauses', but words in their own right, used 'to announce that [speakers] are initiating what they expect to be minor or major delay' in their speech (Clark and Fox Tree, 2002: 103). However, whilst these words may indicate and, more importantly, be *understood* to indicate the expectation of these pauses in *spoken* communication, the same is not necessarily true of *written* communication. As Hammersley notes, 'in ordinary interaction, most of the time, we are not concerned with the particular words used, or the pauses, for their own sake, but rather with understanding what is being said, what its implications are, what responses need to be made, and so on (2010: 560). In some cases, such as ours, excessive precision and detail can obscure more analytically important elements of the data (Bogen, 1992). The mode of transcription should, as such, 'reflect and be sensitive to an investigator's general theoretical model of relations between meaning and speech [and] selectively focus on aspects of speech that bear directly on the specific aims of the study' (Mishler, 1991: 49, see also; Ochs, 1979; Hammersley, 2010).

Spoken and written language are and should be used in different contexts. The process by which naturally 'broken' or 'disfluent' speech is rendered into 'clean' transcripts involves a process of 'intralingual' translation, from one medium (speech) to another (text). When we hear speech, the 'primary' message is the one we hear, and this primary message is (usually) that which the interlocutor wishes to express to us. That which Clark and Fox Tree refer to as the 'collateral message', however, is not heard in spoken discourse as content but instead usually perceived (rightly) as an *explanation* for the structure of the sentence; the use of uhm or uh is to indicate an *expected* pause through the verbalised collateral message which we understand as saying 'I'm still formulating the next part of my speech don't interrupt me', for example, or 'the thing I just said was wrong, I'm going to correct it'. Studies show that, rather than acting as disruptions to speech acts, 'the information available in disfluencies can help listeners compensate for disruptions and delays in spontaneous utterances' (Brennan and Schober, 2001: 274). However, these collateral messages do not tend to tell us much about the substance or meaning of speech, only about the speech act itself: they are used by speakers to comment on the *performance* of speech.(Clark and Fox Tree, 2002) When the content or substance of speech constitutes the purpose for including an extract of transcribed material in disseminated work, these collateral messages cease to be understood in this way.